\documentclass[psfig]{aa}
\input psfig.tex

\begin{document}                                                                

\twocolumn

%-------------------- own definitions -----------------------                   
% ------------------------------------------------                              
%  \thesaurus{03         % A&A Section 6: Form. struct. and evolut. of stars    
%             (03.11.1)}  % Cosmogony,                                         

   \title{Probing Turbulence in the Coma Galaxy Cluster
  \thanks{
          Based on observations with XMM-Newton, an ESA Science Mission
          with instruments and contributions directly funded by ESA Member
          States and the USA (NASA)}}

   \titlerunning{Probing Turbulence}

   \author{P. Schuecker$^{(1)}$, A. Finoguenov$^{(1)}$, 
   F. Miniati$^{(2)}$, H. B\"ohringer$^{(1)}$ and U.G. Briel$^{(1)}$}
                                       
   \authorrunning{Schuecker et al.}

   \offprints{Peter Schuecker\\ peters@mpe.mpg.de}

   \institute{
    $^{(1)}$ Max-Planck-Institut f\"ur extraterrestrische Physik,
             Giessenbachstra{\ss}e, 85741 Garching, Germany\\
    $^{(2)}$ Max-Planck-Institut f\"ur Astrophysik,
             Karl-Schwarzschild-Str., 85741 Garching, Germany\\}

   \date{Received  ; accepted }                         
   
   \markboth{Probing Turbulence}{}

\abstract{Spatially-resolved gas pressure maps of the Coma galaxy
cluster are obtained from a mosaic of XMM-Newton observations in the
scale range between a resolution of 20\,kpc and an extent of
2.8\,Mpc. A Fourier analysis of the data reveals the presence of a
scale-invariant pressure fluctuation spectrum in the range between 40
and 90\,kpc and is found to be well described by a projected
Kolmogorov/Oboukhov-type turbulence spectrum. Deprojection and
integration of the spectrum yields the lower limit of $\sim 10$
percent of the total intracluster medium pressure in turbulent
form. The results also provide observational constraints on the
viscosity of the gas. }

\maketitle

\section{Introduction}\label{INTRO}

In hierarchical structure formation scenarios clusters grow via
accretion and merging of smaller subclumps. Gas accreting onto
clusters of galaxies has bulk velocities of about
$v=1900\,(T/6.7\,{\rm keV})^{0.52}\,{\rm km}\,{\rm s}^{-1}$ at 1\,Mpc
(e.g. Miniati et al. 2000), where $T$ is the mean X-ray temperature of
the intracluster medium (ICM). This velocity is comparable to the
expected sound speed of 1000-1500\,km/s of the ICM. Accretion flows
through filaments and sheets are highly asymmetric and produce complex
patterns which can survive for long time-scales in the ICM (Miniati et
al. 2000).  Simulations by Norman \& Bryan (1999) predict that the
turbulent pressure in the ICM can account for up to 20\% of the
thermal pressure.  We thus expect some measurable effects of
turbulence in the ICM of clusters of galaxies.

Concerning X-ray data, Inogamov \& Sunyaev (2003) propose a study of
spectral line profiles as a useful diagnostic tool of turbulent flows
in the ICM which could be measured with the future ASTRO-E2
satellite. Vogt \& En{\ss}lin (2003) propose the application of
Faraday Rotation measures to test turbulence in the ICM, and claim
that for a few clusters a Kolmogorov spectrum seems to be plausible.

In the present investigation we show that turbulence in the ICM can be
probed directly with pressure maps provided by the XMM-Newton
satellite as a result of its high sensitivity and excellent spectral
capabilities.

Section\,\ref{PHENOMEN} summarizes the basic phenomena related to
turbulent flows. In Sect.\,\ref{EFFECTS} we give a simple analytic
treatment of projection effects introduced through observation. In
Sect.\,\ref{MAPS} we present the observational data and describe how
our X-ray temperature and pressure maps are constructed. Based on the
direct comparsion of local temperature and density measurements, we
give in Sect.\,\ref{GENERAL} some arguments that their observed
fluctuations appear to be almost adiabatic. The same statistical
analysis also suggests the absence of pronounced contact
discontinuities and strong shocks. These observations provide a
baseline consistent with the presence of a turbulent flow. Therefore,
we study in detail in Sect.\,\ref{STRUCT} the measured pressure
spectrum in Fourier-transformed $k$ space and discuss its
interpretation in Sect.\,\ref{DISCUSS}. For all computations a flat
geometry and a Hubble constant of $H_0=50\,{\rm km}\,{\rm
s}^{-1}\,{\rm Mpc}^{-1}$ are used. We assume a distance of $139$\,Mpc
to the Coma cluster so that 1\,arcmin corresponds to about 40\,kpc.

\begin{figure}
\vspace{-0.2cm}
\centerline{\hspace{-0.5cm}
\psfig{figure=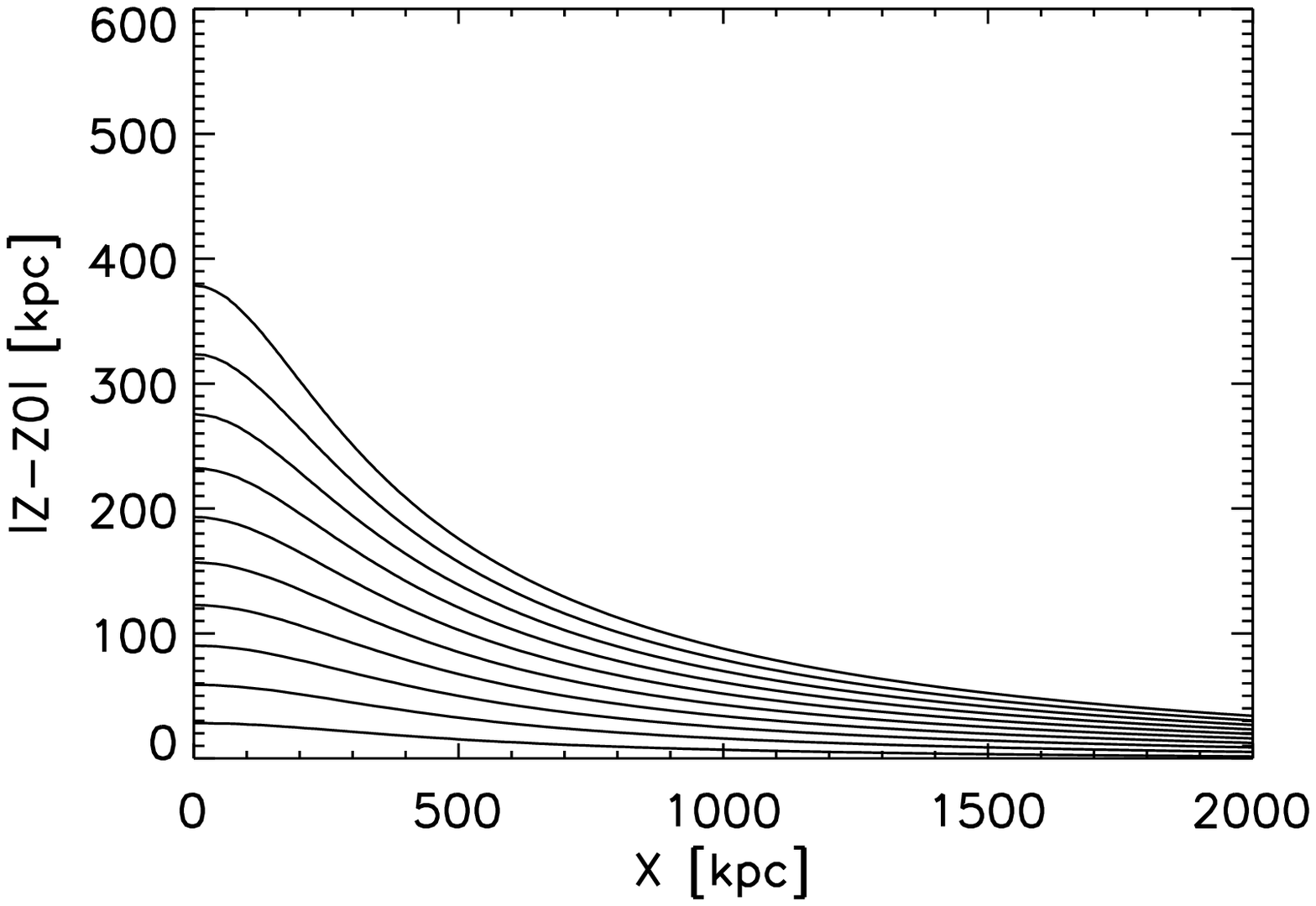,height=6.0cm,width=8.0cm}}
\vspace{0.0cm}
\centerline{\hspace{-0.5cm}
\psfig{figure=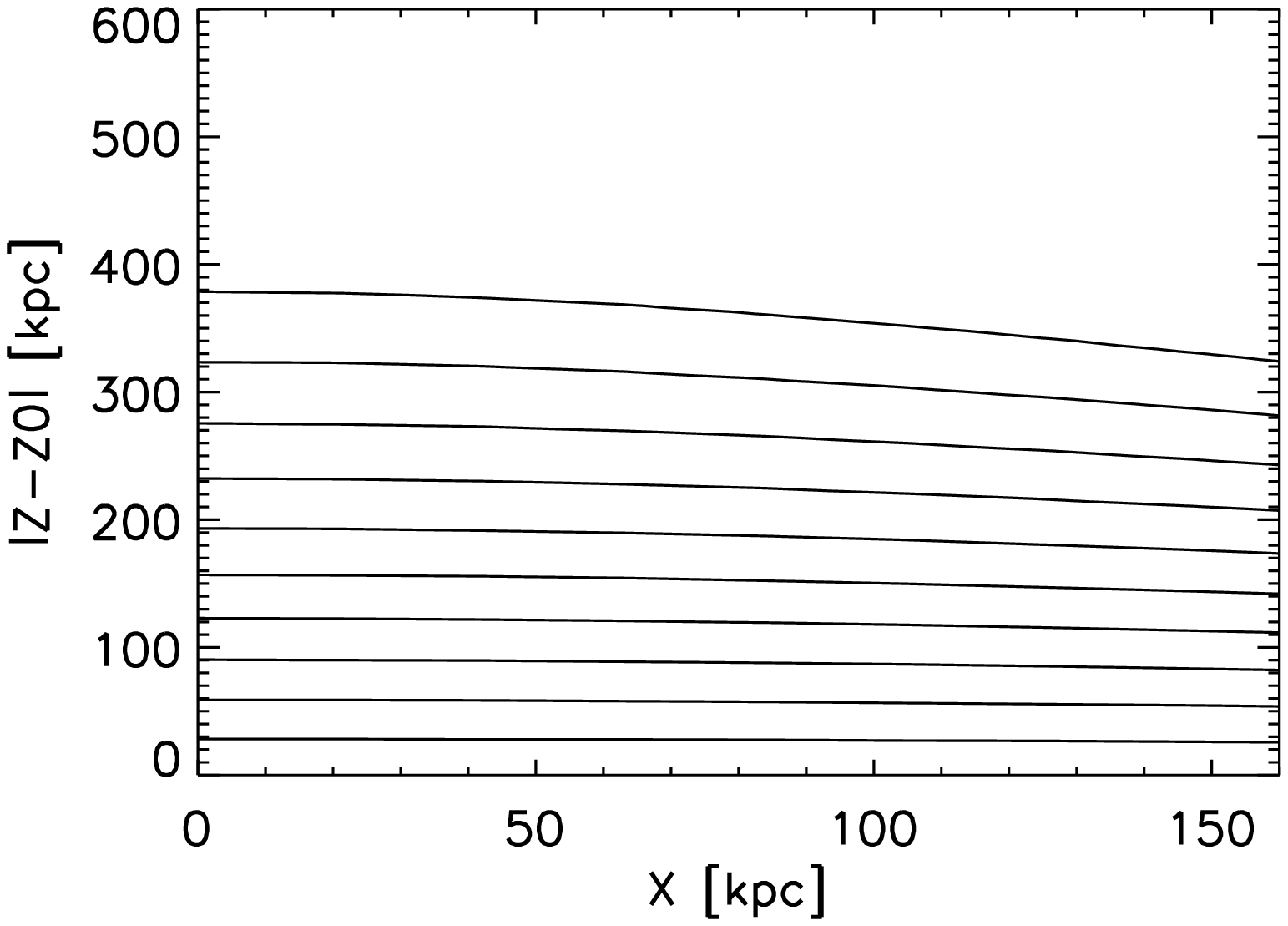,height=6.0cm,width=8.0cm}}
\vspace{-0.0cm}
\caption{\small Angular profile of the weight function ($\beta$
profile squared) integrated along the $z$ direction for various
integration lengths. The lowest/highest contours indicate 10/100\%
contributions, and the intermediate lines the contributions in steps
of 10\%. The upper panel shows the contributions on cluster scales and
the lower panel the contributions on the maximum scales of the
observed turbulence structures (see Sect.\,\ref{DISCUSS}). For the
brightness distribution we assume a $\beta$ profile with the parameter
values $\beta=0.75$ and $r_{\rm c}=420$\,kpc.}
\label{FIG_PROF}
\end{figure}

\begin{figure}
\vspace{-0.5cm}
\centerline{\hspace{0.0cm}
\psfig{figure=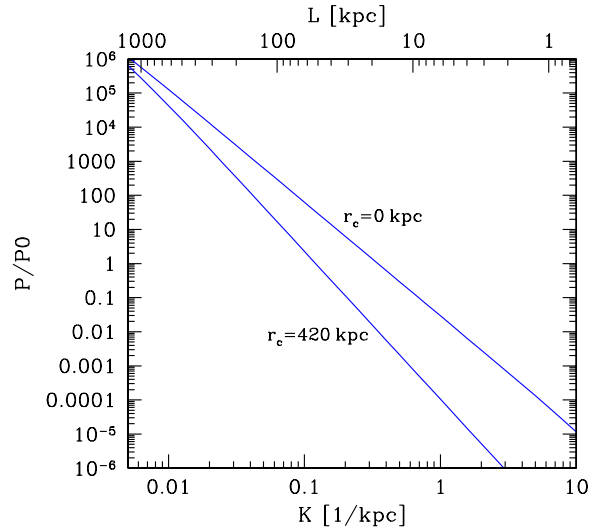,height=8.5cm,width=8.5cm}}
\vspace{-0.45cm}
\caption{\small Pressure power spectra with 
an intrinsic slope of $n=-7/3$ as expected for a Kolmogorov/Oboukhov
turbulence ($r_{\rm c}=0$), and its projection (420\,kpc), as seen
along the $z$ direction through the central plane of a cluster with a
core radius ($r_{\rm c}=420$\,kpc) and slope parameter ($\beta=0.75$)
as measured for the Coma cluster. In order to illustrate the
projection effects over a large scale range, we did not introduce any
characteristic scale which limits the spectra at large and small
scales.}
\label{FIG_MODEL}
\end{figure}

\section{Phenomenology of isotropic turbulence}\label{PHENOMEN}

Traditionally, the phenomenology of isotropic turbulence is based
either on second-order velocity statistics (Kolmogorov 1941) or on
their Fourier-transformed counterparts (Oboukhov 1941). The velocity
energy spectrum
\begin{equation}\label{KOL1}
E_v(k)\,=\,C_{\rm K}\,\epsilon^{2/3}\,k^{-5/3}\,,
\end{equation}
with the non-dimensional Kolmogorov constant $C_{\rm K}$, can be
obtained for the inertial scale range from simple dimensional
arguments (e.g., Lesieur 1997). In Eq.\,(\ref{KOL1}), $E_v(k)$ is the
kinetic energy of fluctuations per unit mass and wavenumber $k$
(physical units ${\rm m}^3{\rm s}^{-2}$), and $\epsilon$, in units of
kinetic energy per unit mass and time (${\rm m}^2{\rm s}^{-3}$), is
the rate of kinetic energy transport from large to small scales.
Eq.~(\ref{KOL1}) fundamentally stems from the assumption that, based
on dimensional arguments, a turbulent eddy on a scale $\lambda
\sim k^{-1}$, decays on a `turnover' timescale 
$\tau=\lambda/v(k)=[k^3E_v(k)]^{-1/2}$ (Oboukhov 1941).  Since the
kinetic energy associated with fluctuations over a scale $k^{-1}$ is
$kE_v(k)$, in steady state regime the rate of energy transport 
across different scales is $\epsilon=kE_v/\tau$ which leads to
Eq.\,(\ref{KOL1}).

\begin{table}
{\bf Tab.\,1.} Spectrum $E_v(k)$ in the inertial range for
different fluids, Mach numbers $\cal{M}$, and magnetic fields $B$.\\
\vspace{-0.5cm}
\begin{center}
\begin{tabular}{lcllc}
\hline
Turbulence & $\cal{M}$ & B-field & $E_v(k)$\\
\hline
$[1]$ &$<1$ & $B=0$  & $k^{-5/3}$\\
$[2]$ &$>1$ & $B=0$  & $k^{-6/3}$\\
$[3]$ &$<1$ & $B\ne0$& $k^{-5/3}$\\
$[4]$ &$>1$ & $B\ne0$& $k^{-3/2\ldots-9/3}$\\
\hline
\end{tabular}
\end{center}
[1] Kolmogorov (1941), Oboukhov 1941), [2] Burgers (1974), [3]
Goldreich \& Sridhar (1995), [4] Cho \& Lazarian (2002), Vestuto et
al. (2003).
\end{table}

In the inertial scale range where Eq.\,(\ref{KOL1}) applies,
turbulence develops without being affected by boundaries, external
forces, or viscosity. Here, the fluctuating quantities are assumed to
be statistically invariant under translation (homogeneity) and
rotation (isotropy). Tab.\,1 summarizes some of the results from
theoretical studies and numerical simulations, which suggest that
Kolmogorov/Oboukhov-like spectra emerge in an inertial scale range
under quite general conditions.

While all these studies are based on the analysis of velocity
fluctuations, Oboukhov (1949) and Batchelor (1951) showed that gas
pressure fluctuations also obey a scaling law (e.g., Lesieur 1997,
Chap.\,VI),
\begin{equation}\label{OB1}
E_P(k)\,=\,C_P\,\epsilon^{4/3}\,k^{-7/3}\,,
\end{equation}
where $C_P$ is a non-dimensional constant, and $E_{\rm P}(k)$ has the
units of kinetic energy per unit mass squared and is normalized to
unit wavenumber $k$ (physical units ${\rm m}^5{\rm s}^{-4}$). We note
that the slope of the spectrum of the pressure is steeper than the
spectrum of the velocity ($P\sim v^2$). As for the velocity spectrum,
we also expect that in general the exact slope of the pressure
spectrum depends on whether or not the fluid is supersonic, and
whether or not magnetic fields are present. The pressure spectrum thus
appears as an excellent and powerful diagnostic tool of turbulent ICM
flows. In addition, this approach appears quite attractive because
pressure fluctuations can already be measured from high resolution
X-ray data.

Establishing the presence of a turbulent ICM implies testing whether
or not there is a scale-free inertial range in the pressure spectrum
with a slope similar to the Kolmogorov/Oboukhov case. It is of further
interest to measure the location of certain characteristic scales,
such as the spectral break which corresponds to the scale where the
kinetic energy is initially injected into the ICM, as well as the
smallest scale where the corresponding energy is finally dissipated
into the ICM. It is, however, not yet clear whether there is also
energy dissipation within the inertial range caused by the development
of randomly distributed weak shocks (Burgers turbulence, Tab.\,1).

\begin{figure*}
\vspace{-0.0cm}
\centerline{\hspace{0.0cm}
\psfig{figure=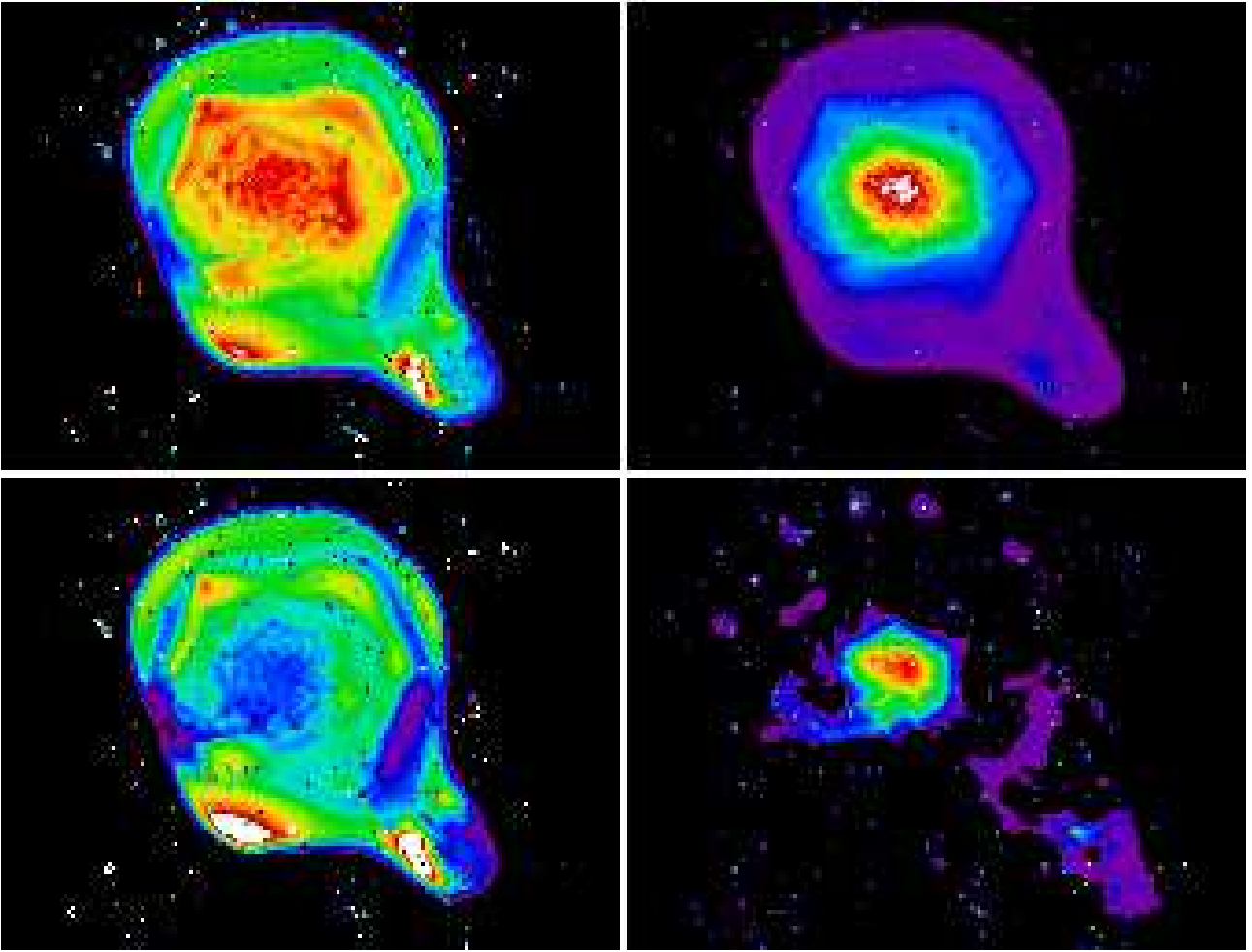,height=13.0cm,width=17.0cm}}
\vspace{-0.0cm}
\caption{\small Projected temperature map (upper left),
pressure map (upper right), entropy map (lower left) and image
substructure (residual) map as revealed by small-scales in the wavelet
decomposition (lower right). The maps are obtained from spectral
hardness ratios and surface brightness data and smoothed with a
wavelet filter of the Coma cluster. Each map covers an area of
$93\times 93\,{\rm arcmin}^2$.}
\label{FIG_WL}
\end{figure*}

\begin{figure*}
\vspace{-0.0cm}
\centerline{\hspace{0.0cm}
\psfig{figure=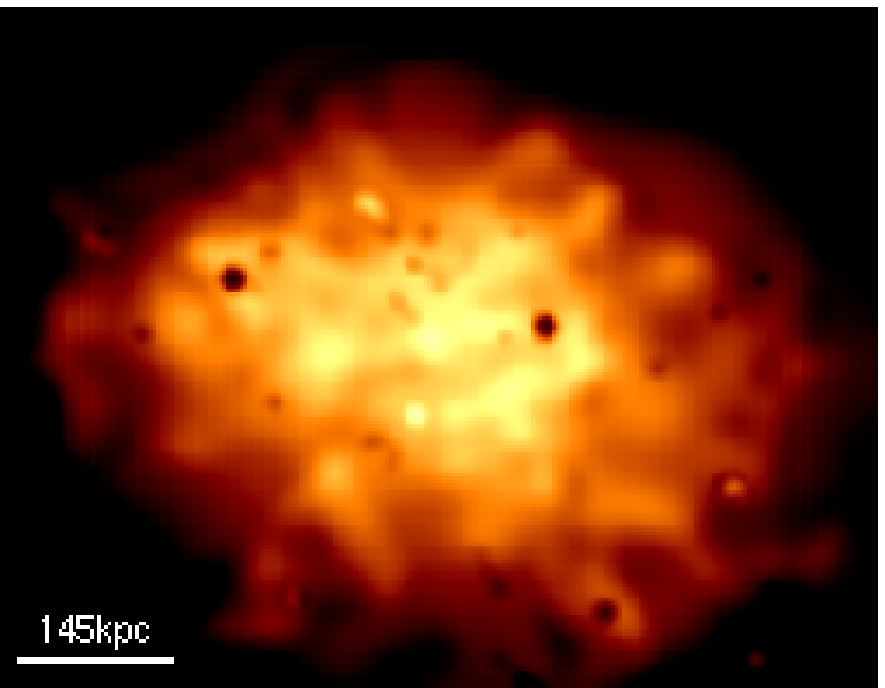,height=14.5cm,width=17.5cm}}
\vspace{-0.0cm}
\caption{\small Detailed view of the projected pressure distribution of the
central region of the Coma cluster. The 145\,kpc scale corresponds to
the largest size of the turbulent eddies indicated by the pressure
spectrum (Sect.\,\ref{DISCUSS}). The smallest turbulent eddies have
scales of around 20\,kpc. On smaller scales the number of photons used
for the spectral analysis is too low for reliable pressure
measurements.}
\label{FIG_PWLZ}
\end{figure*}

\section{Projection effects}\label{EFFECTS}

X-ray observations measure, after filtering and normalization
(Sect.\,\ref{STRUCT}), a projection of the actual three-dimensional
pressure fluctuations, $\delta P/P(\vec{r})$, on the two-dimensional
celestial sphere. Note that the application of normalized instead of
absolute quantities modifies the physical units of the structure
function and power spectrum introduced in Sect.\,\ref{PHENOMEN}. Our
analysis concentrates on scales which are small compared to the
cluster core radius $r_{\rm c}$. As larger scales of order $r_{\rm c}$
are approached, the global cluster profile starts to be probed. This
will be taken into account in the treatment (see
Sect.\,\ref{STRUCT}). Thus, we start by decomposing the fluctuations
into Fourier modes,
\begin{equation}\label{E10}
\frac{\delta P}{P}(\vec{r})\,=\,
\frac{1}{(2\pi)^3}\,\int\,d^3\vec{k}\,\frac{\delta P}{P}(\vec{k})\,
e^{-i\vec{k}\cdot\vec{r}}\,.
\end{equation}
The small-angle approximation allows us to regard the cluster region
as composed of coplanar layers, with a pressure distribution assumed
to be homogeneous and isotropic within each layer.  We deliberately
choose an X-ray energy band which is almost free of the temperature
dependence of emissivity (see Sect.\,\ref{MAPS}). A two-dimensional
pressure field can thus be constructed using $\vec{r}=(\vec{R},z)$,
\begin{equation}\label{E20}
\frac{\delta P}{P}(\vec{R})\,\sim\,
\int\,dz\,\frac{\delta P}{P}(\vec{r})\,\left[1+\frac{(z-z_0)^2}{r_{\rm
c}^2}\right]^{-3\beta}\,,
\end{equation}
which weights the pressure field of each layer with its emissivity, as
obtained from a $\beta$ model of the cluster gas density (squared). In
Eq.\,(\ref{E20}), $z_0$ is the distance between the observer and the
cluster centre along the line-of-sight (LOS), and $\vec{R}$ a
two-dimensional vector on the sky.

An illustration of the weighting scheme is shown for the Coma cluster
in Fig.\,\ref{FIG_PROF}. Here, the contours give the percentage of
surface brightness contributed by ICM gas within a distance $|z-z_0|$
from the plane through the cluster centre and perpendicular to the $z$
direction, as obtained by the integral of the squared $\beta$ profile
along the $z$ direction. A comparison of the upper and lower panels of
Fig.\,\ref{FIG_PROF} reveals that the angular dependence of the
general profile of the Coma cluster is imprinted on large scales, of
order $>200$\,kpc, whereas on smaller scales the structure of Coma
appears quite homogeneous in each layer, so that only a unique profile
along the $z$ direction is seen. In the inner region, weighting can
thus be approximated by taking into account only the variation of the
density along the $z$ direction and neglecting the angular dependence
as assumed in Eq.\,(\ref{E20}). For clusters with small core radii the
approximation is less valid and the projection should be performed
numerically.

A direct consequence of Eq.\,(\ref{E20}) or more complicated
projection schemes is the invariance of relations between fluctuating
quantities under geometric projections. This is illustrated for the
adiabatic relation between temperature and density, $T\sim
n^{\gamma-1}$, used in Sect.\,\ref{GENERAL} to classify the
fluctuations. Its differential version
\begin{equation}\label{E21}
\frac{\delta T}{T}(\vec{r})\,=\,(\gamma\,-\,1)\,\frac{\delta n}{n}(\vec{r})\,.
\end{equation}
defines the (adiabatic) fluctuations.  The projected temperature and
density fluctuations are given by Eq.\,(\ref{E20}) replacing $P$ with
$T$ and $n$. We further replace $[1+(z-z_0)^2/r_{\rm c}^2]^{-3\beta}$
by $W(\vec{r})$, i.e., a general weighting function which describes
the geometric projection process. If the weighting $W(\vec{r})$ were
the same for $T$ and $n$, then we could write
\begin{eqnarray}\label{E22}
\frac{\delta T}{T}(\vec{R})\,&=&\,{\rm const}\,\int\,dz\,
\frac{\delta T}{T}(\vec{r})\,W(\vec{r})\nonumber\\
&=&\,(\gamma-1)\,{\rm const}\int\, dz \,
\frac{\delta n}{n}(\vec{r})\,W(\vec{r})\nonumber\\
&=&\,(\gamma-1)\,\frac{\delta n}{n}(\vec{R})\nonumber\,.
\end{eqnarray}
However, two-dimensional (projected) temperature maps, $T(\vec{R})$,
resulting from X-ray observations, are $n^2(\vec{r})$-weighted
averages along the LOS of the three-dimensional field $T(\vec{r})$. On
the contrary, projected squared densities, $n^2(\vec{R})$, are
obtained by a simple geometric mean without any weighting. For this
more realistic case we thus have
\begin{equation}\label{E23}
\frac{\delta T(\vec{R})}{T(\vec{R})}=\frac{\int\, dz\,\delta
T(\vec{r})\, n^2(\vec{r})}{\int\, dz\, T(\vec{r})\, n^2(\vec{r})}
=\frac{\gamma-1}{2}\frac{\int\, dz\, T(\vec{r})\,\delta
n^2(\vec{r})}{\int\, dz\, T(\vec{r})\, n^2(\vec{r})}\,,
\end{equation}
where $\delta n^2=2n\,\delta n$ and Eq.\,(\ref{E21}) have been
used. For the assumed central core region of the cluster we now have
the advantageous situation that $T(\vec{r})$ assumes the role of an
almost constant weighting function for the density, and we can show
that retaining only first order terms, Eq.\,(\ref{E23}) simplifies to
\begin{equation}\label{E24}
\frac{\delta
T(\vec{R})}{T(\vec{R})}\,\approx\,\frac{\gamma-1}{2}\frac{\int\,dz\,\delta
n^2(\vec{r})}{\int\,dz\,n^2(\vec{r})}\,=\,\frac{\gamma-1}{2}\,
\frac{\delta n^2(\vec{R})}{n^2(\vec{R})}\,.
\end{equation}
Eq.\,(\ref{E24}) is the observational counterpart of Eq.\,(\ref{E21})
and will be used in Sect.\,\ref{GENERAL} to constrain possible types
of pressure fluctuations measured in the centre of the Coma cluster.

After cross-correlating projected fluctuation measures, we now proceed
with projecting pressure spectra. We still assume that the central
cluster region, to which our analyses are restricted, is small enough
to be approximated by a set of coplanar, homogeneous, and isotropic
layers, yielding a unique pressure profile along the $z$ direction. 
We want to take advantage of the invariance property of a Gaussian
profile under Fourier transformation by replacing the $\beta$ model by
a Gaussian with variance $r_{\rm c}^2/3\beta$. The approximation is
better than 5\% for $r<r_{\rm c}$ and allows us to regard the
convolution of the pressure field with the gas density profile in
Eq.\,(\ref{E20}) along the $z$ direction as a transformation $\delta
P/P(k)\rightarrow\delta P/P(k)\exp{(-k_z^2r_{\rm c}^2/6\beta)}$. With
the standard relations between two and three-dimensional power spectra
(e.g., Peacock 1999, Sect.\,18.1) we obtain the simple expression
\begin{equation}\label{E30}
{\cal P}_{\rm 2D}(K)\,=\,\frac{1}{\pi}\,\int_0^\infty\,dk_z\,{\cal
P}_{\rm 3D}\left(\sqrt{k_z^2+K^2}\right)\,
\exp{\left(-\frac{k_z^2\,r_{\rm c}^2}{3\beta}\right)}\,,
\end{equation}
where the integration extends over the wavenumber $k_z$ along the $z$
direction. The wavenumbers in two and in three dimensions are $K$ and
$k$, respectively. Note that $K=2\pi/\Theta$, with $\Theta$ being the
angular scale of the projected fluctuation, has the physical units
${\rm rad}^{-1}$, but can be readily transformed to metric scales
using the distance to the cluster. Eq.\,(\ref{E30}) shows that adding
three-dimensional fluctuations along the $z$ direction leads to an
exponentially damped two-dimensional spectrum. Damping is larger for
smaller modes because relatively more fluctuations are added along the
$z$ direction.

\begin{figure*}
\vspace{0.2cm}
\centerline{\hspace{0.0cm}
\psfig{figure=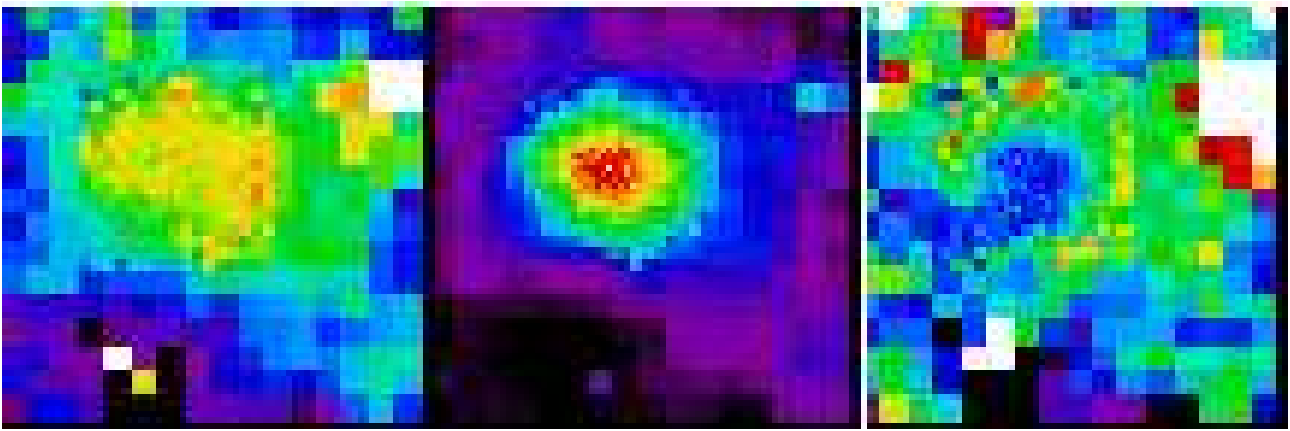,width=17.0cm}}
\vspace{-0.0cm}
\caption{\small Nested grids of temperature (left panel), 
pressure (middle panel), and entropy (right panel) measurements. Each
map covers an area of $69.3\times 69.3\,{\rm arcmin}^2$.}
\label{FIG_PRO}
\end{figure*}

The relations between power spectra and pressure spectra in two and
three-dimensions can be obtained from the condition that the sum over
both statistics must give the same total energy,
\begin{equation}\label{E40}
\frac{1}{(2\pi)^2}\,\int_0^\infty\,2\pi\,K\,dK\,{\cal P}_{\rm
2D}(K)\,=\,\int_0^\infty\,E_{\rm 2D}(K)\,dK\,,
\end{equation}
\begin{equation}\label{E50}
\frac{1}{(2\pi)^3}\,\int_0^\infty\,4\pi\,k^2\,dk\,{\cal P}_{\rm
3D}(k)\,=\,\int_0^\infty\,E_{\rm 3D}(k)\,dk\,,
\end{equation}
which yields the following conversions:
\begin{eqnarray}\label{E60}
{\cal P}_{\rm 2D}(K)\,&=\,2\,\pi\,K^{-1} &\,E_{\rm 2D}(K)\,,\\ {\cal
P}_{\rm 3D}(k)\,&=\,2\,\pi^2\,k^{-2}&\,E_{\rm 3D}(k)\,.
\end{eqnarray}
This gives the projection of the pressure spectrum
\begin{equation}\label{E70}
E_{\rm 2D}(K)=K\int_0^\infty\frac{dk_z}{k_z^2+K^2}\,
E_{\rm 3D}(\sqrt{k_z^2+K^2})\,e^{-\frac{k_z^2r_{\rm c}^2}{3\beta}}\,,
\end{equation}
where $E_{\rm 3D}(\cdot)$ can be identified with the spectrum
Eq.\,(\ref{OB1}) after including appropriate upper and lower
integration limits. Note that the decline of $E_{\rm 3D}(k)$ and thus
of ${\cal P}_{\rm 3D}(k)$ outside the inertial range may not be sharp,
and instead better described by smooth characteristic shape functions
(see, e.g., Fritsch et al. 1980). Fig.\,\ref{FIG_MODEL} gives a
quantitative impression of the projection effects on a power spectrum
computed with Eq.\,(\ref{E30}) for the Coma cluster with distance
139\,Mpc, $\beta=0.75$, and core radius $r_{\rm c}=420$\,kpc. For
illustration, no $K$ limits of the inertial range are introduced.  For
comparison we also plot the $k_zr_{\rm c}\ll 1$ case where no
corrections for projection are necessary. The intrinsic slope of the
pressure spectrum is $n=-7/3$. Note that the power spectrum of the
pressure fluctuations is damped at 60\,kpc by a factor of about
28. The observation of pressure fluctuations along the $z$ direction
through the cluster thus damps their amplitudes by a factor
$\sqrt{28}\approx 5.3$. However, cluster-wide fluctuations on
Mpc-scales would appear almost undamped.

Further observational effects are related to the measurement errors of
the pressure and the global pressure profile and can be illustrated
best with the observed spectrum of the cluster pressure distribution
(Sect.\,\ref{STRUCT}).

\section{The Coma mosaic}\label{MAPS}

\begin{table*}[t]
{\bf Tab.\,2.} Summary of XMM-Newton observations \\
\vspace{-0.1cm}
\begin{center}
\begin{tabular}{llcccccc}
\hline
  & Name of     &          &           &  \multicolumn{2}{c}{pn-Camera
observing times (ksec)} & Orbit & Frametime\\
\multicolumn{1}{c}{Date} & Observation & RA(2000) & DEC(2000) &
planned     & \multicolumn{1}{c}{effective} &  & msec\\
\hline
2000 May 29      & Coma center & 12 59 46.7 & 27 57 00 & 15.0 & 12.8 & 86 &199 \\
2000 June 21     & Coma 1      & 12 56 47.7 & 27 24 07 & 25.0 & 21.1 & 98 & 73 \\
2000 June 11     & Coma 2      & 12 57 42.5 & 27 43 38 & 25.0 & 43.8 & 93 &199 \\
2000 June 27     & Coma 3      & 12 58 32.2 & 27 24 12 & 25.0 & 11.0 &101 & 73 \\ 
2000 June 23     & Coma 4      & 13 00 04.6 & 27 31 24 & 25.0 &  5.4 & 99 &199 \\
2000 May 29      & Coma 5      & 12 59 27.5 & 27 46 53 & 20.0 &  9.9 & 86 &199 \\
2000 June 12     & Coma 6      & 12 58 50.0 & 27 58 52 & 20.0 & 12.4 & 93 &199 \\
2000 Dec 10      & Coma 7      & 12 57 27.7 & 28 08 41 & 25.0 & 18.6 &184 &199 \\ 
2000 Dec 10/11   & Coma 8      & 13 01 25.6 & 27 43 53 & 26.0 & 14.3 &184 &199 \\  
2000 June 11/12  & Coma 9      & 13 00 32.7 & 27 56 59 & 20.0 & 14.7 & 93 &199 \\
2000 June 22     & Coma 10     & 12 59 38.4 & 28 07 40 & 20.0 & 15.4 & 98 &199 \\
2000 June 24     & Coma 11     & 12 58 36.5 & 28 23 56 & 25.0 & 11.5 & 99 &199 \\
2002 June 5/6    & Coma 12     & 13 01 50.2 & 28 09 28 & 25.0 &  9.6 &456 &199 \\ 
2002 June 7/8    & Coma 13     & 13 00 36.5 & 28 25 15 & 25.0 & 20.0 &457 &199 \\
2000 June 22     & Coma 0 (bkg)& 13 01 37.0 & 27 19 52 & 30.0 & 12.8 & 98 &199 \\
2001 Dec 4/5     & Coma cal    & 12 59 46.6 & 27 57 00 & 25.0 & 17.4 &364 &73 \\ 
\hline
\end{tabular}
\end{center}
\end{table*}

In this paper we use the performance verification observations of the
Coma cluster obtained with the EPIC-pn instrument on board of
XMM-Newton (Jansen et al. 2001). Previous reports of these
observations were given by Briel et al. (2001), Arnaud et al. (2001),
Neumann et al. (2001, 2003) and Finoguenov et al. (2004a, for
point-like sources). This work includes all datasets obtained to date,
as described in detail by Finoguenov et al. (2004a, see also
Tab.\,2). While most of the pointings have been obtained in Extended
Full Frame Mode (Frametime 199\,msec in Tab.\,2), three observations
were conducted in Full Frame Mode (Frametime 73\,msec).

All observations have been reprocessed using the latest version of the
XMM reduction pipeline (XMMSAS 5.4.1), which yields an astrometry to
better than 1 arcsec. Although the Coma data are public, for some
(Coma-10 and Coma-0) of the early observations of the performance
verification phase no complete Observational Data Files (ODF) had been
produced by standard processing, and a special preprocessing (XMMSAS
task {\it odffix}) was done on such pn exposures at MPE by Michael
Freyberg. As a result, a few Coma pointings are not yet publicly
available, which precluded us from using MOS data. The vignetting
correction, crucial for obtaining reliable source characteristics over
a wide region, is performed using the latest calibration (Lumb et
al. 2003). Two pointings at the Coma centre were used in that
calibration, in a way requiring that the same sky pixels yield the
same flux between the two observations. The level of the emission was
not used in the calibration, so it could be analyzed further. The RMS
fluctuations of the comparison of two Coma fields is within 2\%, which
will affect the apparent pressure fluctuations studied here on the 1\%
level, much lower than the observed 10\% amplitudes.

The images were extracted separately for each pointing, along with the
corresponding exposure maps. We select pn events with ${\rm
PATTERN}<5$ and $({\rm FLAG}
\& {\rm 0xc3b0809}) = 0$, which in addition to ${\rm FLAG}=0$ events 
includes events in the rows close to gaps and bad pixels; however, it
excludes the columns with offset energy. This event selection results
in a better spatial coverage of the cluster, but at a somewhat
compromised energy resolution, which is sufficient for the broad-band
imaging.

Our final results are derived from a spectral analysis where only the
${\rm FLAG}=0$ events were retained. For background subtraction we
used the similarly screened and selected events from the background
accumulation of Andrew Read (Read \& Ponman 2003) and also subtracted
out-of-time events as a background, using products from the SAS task
{\it epchain}. This subtraction is important as some pointings (see
FrameTime 73\,ms in Tab.\,2) are performed in the Full Frame Mode.

To provide an overview of the structure of the ICM of the Coma
cluster, we show in Figs.\,\ref{FIG_WL} and \ref{FIG_PWLZ} the
temperature, pressure, and entropy maps, as well as maps of the
small-scale surface brightness structure. These maps use hardness
ratios in the 0.8--2\,keV and 2--7.5\,keV bands, calibrated for the
measurement of temperature, as a substitute for the temperature
determined directly from the spectral analysis. The projected entropy
($S$) and projected pressure ($P$) maps are derived from the projected
temperature, $T$, and surface brightness, $\Sigma$, through the
relations $S=T\Sigma^{-1/3}$ and $P=T\Sigma^{1/2}$, respectively. The
maps are constructed from composite wavelet filtered images to
suppress the large scale background. The details of the analysis based
on surface brightness and hardness ratio maps and the rationale of the
use of wavelet filtering is described in detail in previous
publications (Briel et al. 2003, Finoguenov et al. 2004b, and Henry et
al. 2004).

The substructures seen in these maps suggest turbulent-like
fluctuations. The pressure maps are of special importance because they
clearly show fluctuations which are not contaminated by contact
discontinuities (see Sect.\,\ref{GENERAL}). For a quantitative study
of the significance of these fluctuations we thus performed direct
fits to the spectral X-ray data.

In Fig.\,\ref{FIG_PRO} we show the temperature, entropy and pressure
maps, based on the temperature and emission measure obtained through
direct spectral fitting, using several grids to define the region of
spectral extraction. Only the $16\times16$ grids with a pixel size
$40\times 40\,{\rm arcsec}^2$, $120\times 120\,{\rm arcsec}^2$,
$260\times 260\,{\rm arcsec}^2$ are shown. This figure also
illustrates the relative positioning of the grids. Each binning
involves a mixing of various spectral components. Therefore, a
decision has to be made on which of the components the spectral
analyses should be performed. We have chosen to put our interest on
the hotter component, and so have used the 1--7.9\,keV energy band for
spectral fitting. Fine grids, with a pixel size of $40\times 40\,{\rm
arcsec}^2$ and lower, located in the central region, do not suffer
that much from temperature mixing, but they do suffer from small
number statistics. So, for those we used the 0.5--7.9 keV band. A
detailed check has shown that for a similar location in the Coma
cluster all grids yield similar temperature estimates, which supports
our choice of energy bands.  The selection of the grid resolution was
performed to yield at least 5000 counts per pixel. The total number of
counts available for the analysis in the Coma observation reaches two
million counts in the 0.5--2 keV band and a similar amount in the
harder band (2--7.9\,keV).

\section{General character of the fluctuations}\label{GENERAL}

\begin{figure}
\vspace{-0.5cm}
\centerline{\hspace{-0.4cm}
\psfig{figure=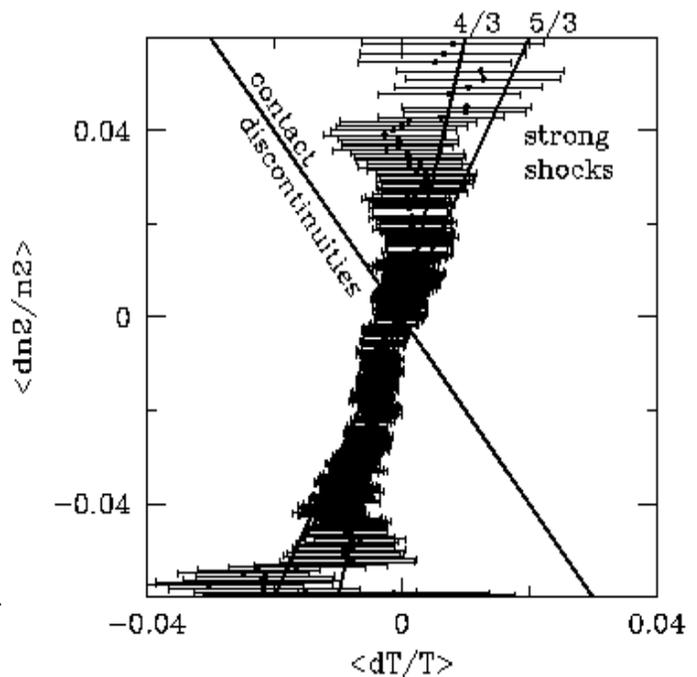,height=9.0cm,width=9.0cm}}
\vspace{-0.25cm}
\caption{\small Correlation between relative fluctuations of density
squared ($n^2$) and temperature ($T$), and their $1\sigma$ errors. The
two thick lines represent the adiabatic exponent $\gamma=5/3$
(monoatomic ideal gas) and $\gamma=4/3$ which gives a good
representation of the data. Due to the sliding window method used,
neighbouring data points are correlated.}
\label{FIG_GAMMA}
\end{figure}

In order to obtain more information about the type of fluctuations
seen in Figs.\,\ref{FIG_WL}--\ref{FIG_PRO}, we performed a
pixel-by-pixel cross-comparison of temperature and density gradients.

Figure\,\ref{FIG_GAMMA} shows the correlation between the gradients of
the projected X-ray temperature and the gradients of the projected
squared gas density as obtained for the $40\times40\,{\rm arcsec}^2$
pixel grid. We concentrate on this specific grid because it is mainly
restricted to the core region of the Coma cluster and has sufficiently
high signal-to-noise X-ray spectra at a comparatively small angular
resolution.

The relative fluctuations are determined for each pixel of the
temperature and density map by averaging the gradients over its four
nearest neighbour pixels.  The averaged gradients are obtained for
density and temperature maps and can be compared in a point-wise
manner. However, the individual fluctuations are large (10\% level)
because they include also the measurement errors. Therefore, an
additional binning with $dn^2/n^2=0.1$ and a continuous sliding of
this bin along the density axis is necessary to see a clear trend. The
error bars are the $1\sigma$ fluctuations of the mean obtained for
each bin.

For the classification of the fluctuations, we show in
Fig.\,\ref{FIG_GAMMA} model expectations obtained with
Eq.\,(\ref{E24}). The line labeled $\gamma=5/3$ corresponds to a
monoatomic ideal gas. For contact discontinuities local pressure
equilibrium leads to $\delta n/n=-\delta T/T$. Apparently, the
$\gamma=4/3$ line gives a better representation than what is expected
for the $\gamma=5/3$ case. This could be due to a contamination by
contact discontinuities. 

Since the gradients are measured on scales of 27\,kpc, which are small
compared to the cluster core radius of $r_{\rm c}=420$\,kpc, we expect
them to be sensitive probes of local substructure fluctuations and not
significantly affected by the global cluster profile. We nevertheless
tested this approximately with Monte Carlo simulations where the
gradients of $n^2$ are computed with an isothermal $\beta$ model of
the Coma cluster and a gas adiabatic equation of state. The gradients
are determined in the same way as the empirical data and added to the
adiabatic density fluctuations. We found that large-scale gradients in
the density field broaden the scaling relation, but without
introducing a bias in the determination of $\gamma$. In fact, we have
verified this for different values of the adiabatic exponent
$\gamma$. 

The observed temperature and density maps are also tested for possible
correlations between $T$ and $n^2$ introduced by the large-scale
distribution of the ICM. The radial profiles are obtained by averaging
temperatures and densities in concentric rings with a width of
50\,kpc. Whereas $n^2$ shows a significant decrease of $27\pm2$
percent between the cluster center and 300\,kpc, the temperature
decreases by only $3\pm2$ percent relative to the central value of
$7.34\pm0.13$\,keV. The observed temperature gradients thus appear
with the same size as the errors and can thus be neglected as a
possible second-order effect. Within this approximation, no
correlations between temperature and density fluctuations are
introduced by the global cluster profile. 

To conclude, Fig.\,\ref{FIG_GAMMA} suggests a positive correlation
between temperature and density gradients which is not related to the
large-scale distribution of the ICM. The gradients occupy different
regions than contact discontinuities and {\it strong} shocks. The data
are actually quite close to the expected adiabatic case.  In order to
find out whether or not such fluctuations are organized as in a
turbulent regime, we study in the following the statistics of the
spatial pressure fluctuations.

\section{Power spectrum of spatial pressure fluctuations}\label{STRUCT}

The first step in our (standard) power spectrum analysis is the
determination of the global pressure profile $\bar{P}(\vec{R})$ from
the observed 2-dimensional pressure map $P(\vec{R})$, in order to get
the residual local pressure fluctuations, $\delta
P/P(\vec{R})=P(\vec{R})/\bar{P}(\vec{R})-1$. The second step is the
determination of the Fourier power spectrum of $\delta P/P(\vec{R})$,
corrected for the errors of the pressure measurements (shot-noise
subtraction), and normalized to unit number of Fourier modes and to
unit area in $K$ space. The resulting projected spectrum ${\cal
P}_{\rm 2D}(K)$ has the physical units ${\rm kpc}^2$. In the following
example, the pressure is measured in a regular grid of $32\times 32$
cells, each with $20\times 20\,{\rm arcsec}^2$. This grid covers the
central core region of Coma up to 431\,kpc and has the fundamental
mode $K=2\pi/\lambda=0.0146\,{\rm kpc}^{-1}$. The results obtained
with the other three grids are given at the end of this section.

\begin{figure}
\vspace{-0.5cm}
\centerline{\hspace{0.0cm}
\psfig{figure=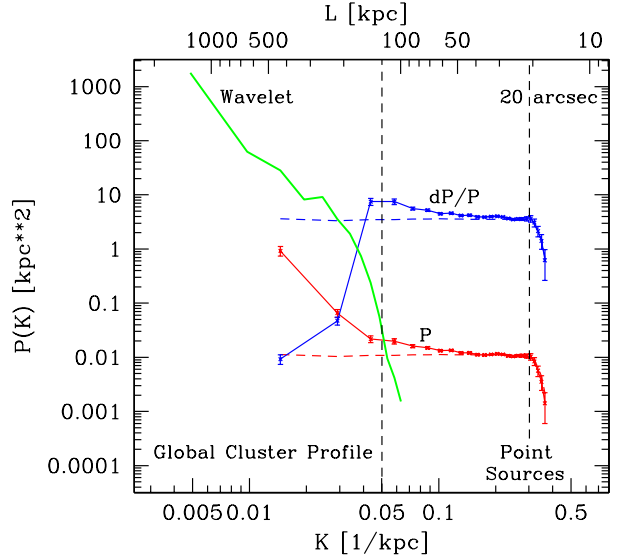,height=9.0cm,width=9.0cm}}
\vspace{-1.25cm}
\caption{\small Projected power spectra of different angular pressure
distributions from the $20\times 20\,{\rm arcsec}^2$ grid. Lower thin
continuous line: raw spectrum including shot-noise, substructure, and
cluster profile (P). Upper thin continuous line: spectrum from the
normalized pressure distribution (dP/P). Thick continuous line:
spectrum of the global cluster pressure profile as determined with the
wavelet transform. Dashed horizontal lines: shot-noise levels computed
from measured pressure errors. Dashed vertical lines: characteristic
scales.}
\label{FIG_ILLUS}
\end{figure}

The global pressure profile $\bar{P}(\vec{R})$ is obtained from a
low-passband Fourier-filter applied to $P(\vec{R})$ with a filter
scale of 150\,kpc, which leaves the global cluster profile above this
scale almost unchanged. To illustrate the effect of the filter, we
show in Fig.\,\ref{FIG_ILLUS} the power spectrum (marked `P') obtained
from a direct Fourier-transformation of $P(\vec{R})$. On scales
between 20 and 40\,kpc, the spectrum has a flat plateau-like
distribution which is determined by the temperature and density errors
(shot-noise, see below). Between 40 and 125\,kpc the spectrum
increases significantly above the shot-noise level. This is the
spectrum of the substructures seen in Figs.\,\ref{FIG_WL} to
\ref{FIG_PRO}. Beyond 125--150\,kpc, the spectrum abruptly
increases due to the global pressure profile of the Coma cluster.

A similar increase is also seen in the spectrum marked `Wavelet' which
is obtained alternatively from a wavelet-filtered pressure map. For
the wavelet decomposition we used the algorithm of Vikhlinin et
al. (1998) and computed the spectrum from the wavelet reconstruction
of the $30\times 30\,{\rm arcsec}^2$ map with the lowest angular
resolution. The wavelet algorithm performs a self-adjusting noise
suppression so that almost no significant shot-noise occurs in the
spectrum of the global pressure profile. The $20\times 20\,{\rm
arcsec}^2$ grid does not cover the complete cluster area and is thus
not optimal for the proper sampling of the global cluster
profile. Therefore, the similarity of the `P' and `Wavelet' spectra is
not very good on large scales. However, grids with larger bin sizes
cover larger scales and give a very good agreement with the `Wavelet'
profile (see below). For the following analyses we thus use the
Fourier low-pass filter with a filter scale of 150\,kpc to determine
$\bar{P}(\vec{R})$ for all four grids.

\begin{figure}
\vspace{-0.5cm}
\centerline{\hspace{0.0cm}
\psfig{figure=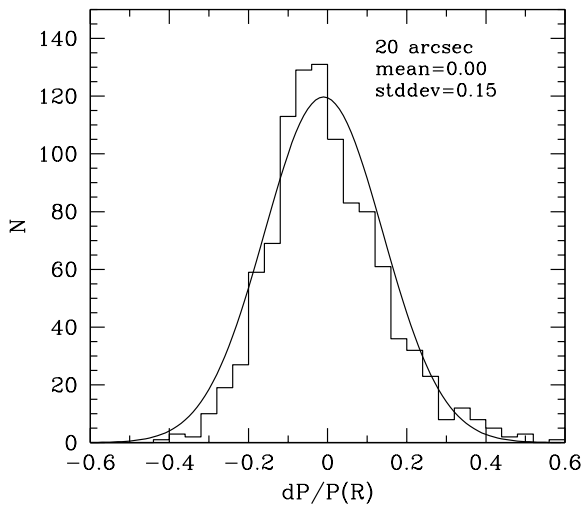,height=9.0cm,width=9.0cm}}
\vspace{-1.00cm}
\caption{\small Comparison of a Gaussian profile with the 
histogram of projected pressure contrasts $\delta
P/P=[P(\vec{R})/\bar{P}]-1$ in configuration space for a pixel size of
$20\times 20\,{\rm arcsec}^2$ which corresponds to $13.5\times
13.5\,{\rm kpc}^2$. The mean over all fluctuation is almost zero and
the $1\sigma$ standard variation $\sigma_{\delta P/P}=0.15$.}
\label{FIG_HISTO}
\end{figure}

The histogram of the resulting $\delta P/P(\vec{R})$ is shown in
Fig.\,\ref{FIG_HISTO}. Their distribution appears quite consistent
with a Gaussian random field (KS-probability of 90\%) with zero mean
and a standard deviation of 15 percent (including shot-noise) on a
pixel scale of 13.5\,kpc. The $\delta P/P(\vec{R})$ field can thus
completely be summarized by a power spectrum. The corresponding power
spectral densities are marked by `dP/P' in Fig.\,\ref{FIG_ILLUS}. The
spectrum shows the expected drop beyond 150\,kpc. A similar drop at
scales below 20\,kpc marks the resolution limit as given by the
pixelation (see `Point Sources'). These two cutoff scales limit the
range of the power spectrum of the substructures.

The effect of temperature and density measurement errors is seen in
the power spectra as an almost scale-independent shot-noise level
which must be subtracted from the `dP/P'-spectrum (dashed horizontal
lines in Fig.\,\ref{FIG_ILLUS}). For the determination of the
shot-noise, we first determine at each grid point the local relative
pressure error $\sigma_P$, using $T$ and its error $\sigma_T$ as well
as $n^2$ and its error $\sigma_{n^2}$, as obtained from the local
X-ray spectral fit. We then draw at each grid point a random value for
the local relative pressure error from a Gaussian distribution with
zero mean and standard deviation $\sigma_P$. After performing this
randomization for all grid points, one realisation of a random map is
generated with fluctuations solely caused by measurement errors.  For
the determination of the shot-noise we averaged the power spectra of
100 random realisations.

The $1\,\sigma$ error bars shown in Fig.\,\ref{FIG_ILLUS} are also
determined from the variances of the spectra obtained from randomized
maps of the measurement errors. The errors are lower limits because
they are obtained from unstructured pressure maps.  Unfortunately,
much larger effort is needed to improve these estimates, for example,
with a set of hydrodynamical cluster simulations.

\begin{figure}
\vspace{-0.5cm}
\centerline{\hspace{0.0cm}
\psfig{figure=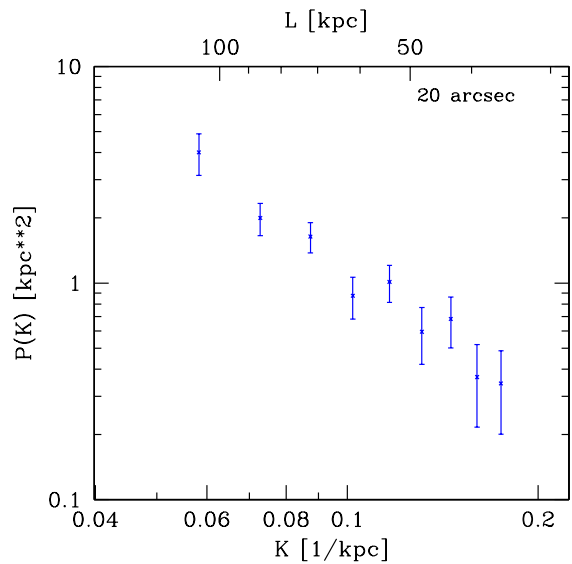,height=9.0cm,width=9.0cm}}
\vspace{-1.00cm}
\caption{\small Projected shot-noise subtracted power spectral
densities, ${\cal P}_{\rm 2D}(K)$, of the pressure fluctuations, and
their $1\sigma$ errors, obtained from the $20\times 20\,{\rm
arcsec}^2$ grid.}
\label{FIG_PKF5}
\end{figure}

Figure\,\ref{FIG_PKF5} shows the power spectrum of $\delta
P/P(\vec{R})$ after shot-noise subtraction. This spectrum can be
compared via Eq.\,(\ref{E30}) with theoretical 3-dimensional power
spectra or structure functions.

\begin{figure}
\vspace{-0.5cm}
\centerline{\hspace{0.0cm}
\psfig{figure=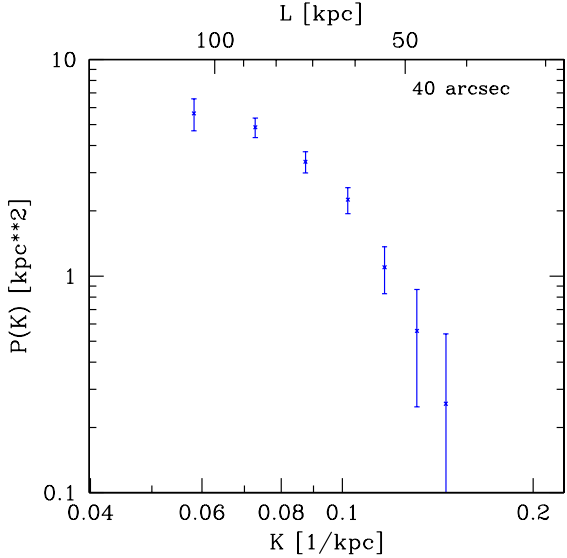,height=9.0cm,width=9.0cm}}
\vspace{-1.00cm}
\caption{\small Projected shotnoise subtracted power spectral
densities, ${\cal P}_{\rm 2D}(K)$, as in Fig.\,\ref{FIG_PKF5} for the
$40\times 40\,{\rm arcsec}^2$ grid.}
\label{FIG_PKF10}
\end{figure}

\begin{figure}
\vspace{-0.5cm}
\centerline{\hspace{0.0cm}
\psfig{figure=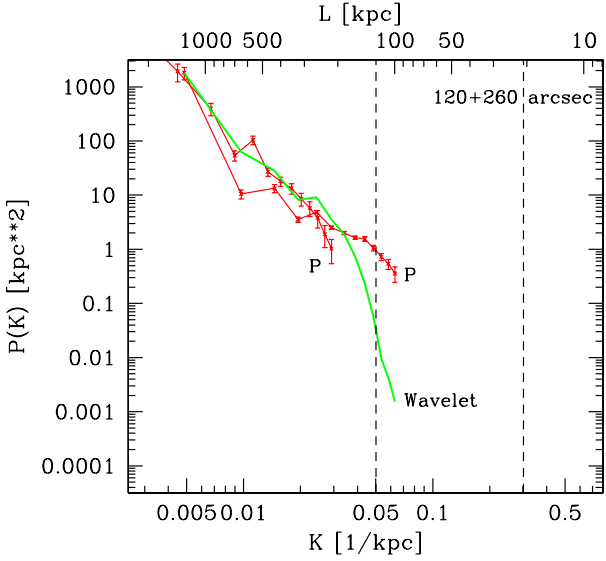,height=9.0cm,width=9.0cm}}
\vspace{-1.00cm}
\caption{\small Projected power spectral
densities as in Fig.\,\ref{FIG_ILLUS} for the
$120\times 260\,{\rm arcsec}^2$ and the $260\times 120\,{\rm
arcsec}^2$ grid. The spectra basically follow the global cluster
pressure profile.}
\label{FIG_PKF3065}
\end{figure}

The same analysis performed on the $40\times 40\,{\rm arcsec}^2$ grid
gives the power spectrum shown in Fig.\,\ref{FIG_PKF10}. The errors
are smaller compared to the results obtained with the $20\times
20\,{\rm arcsec}^2$ grid. We attribute this to the higher
signal-to-noise X-ray spectra obtained with the larger pixels. The
spectral shape appears somewhat more curved and steeper than the
spectrum obtained with the smaller grid. The $120\times 120\,{\rm
arcsec}^2$ and $260\times 260\,{\rm arcsec}^2$ grids mainly sample the
global cluster profile. The resulting `P'-spectra shown in
Fig.\,\ref{FIG_PKF3065} follow the profile obtained with the
wavelet-filtered pressure distribution. These spectra do not have any
significant fluctuation power on scales below 100\,kpc so that we do
not show the corresponding `dP/P'-spectra.

\begin{figure}
\vspace{-0.5cm}
\centerline{\hspace{0.0cm}
\psfig{figure=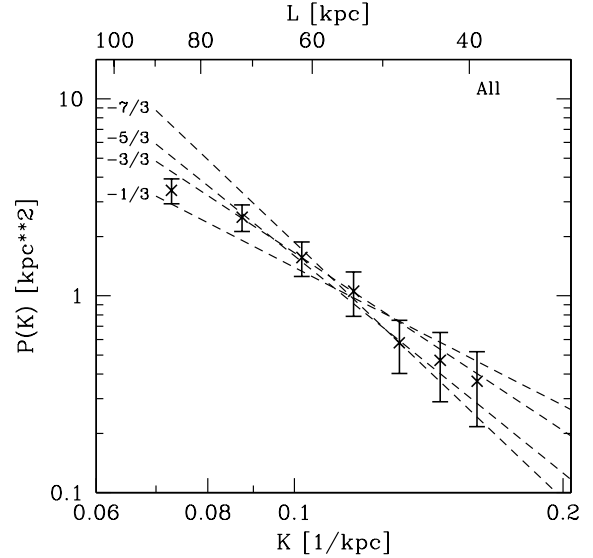,height=9.0cm,width=9.0cm}}
\vspace{-1.00cm}
\caption{\small Observed projected shot-noise-subtracted 
power spectral densities (dots with $1\sigma$ error bars) as obtained
for the $20\times 20\,{\rm arcsec}^2$ and $40\times 40\,{\rm
arcsec}^2$ grids, compared with model predictions (dashed lines).}
\label{FIG_PKFALL}
\end{figure}

The differences seen in the power spectral densities shown in
Figs.\,\ref{FIG_PKF5} and \ref{FIG_PKF10} are caused by the chosen
center, bin size, and total size of the sample grid. This sample
variance can be reduced by averaging the spectral densities measured
at the same $K$-values. The average is meaningful because all spectral
densities -- although determined in $K$ bins with different sizes
(fundamental modes) -- are normalized to the same unit volume of
$1/{\rm kpc}^2$ in $K$ space. The random errors of the spectral
densities which do not refer to sample variance are not reduced
because the two power spectra cannot be regarded as completely
statistically independent. We regard the spectrum shown in
Fig.\,\ref{FIG_PKFALL} as the final result of the power spectrum
analysis.

\section{Discussion}\label{DISCUSS}

The present investigation aims to detect turbulence in the ICM of the
Coma cluster using the pressure spectrum. Under certain
approximations, one also expects a scale-invariant spectrum of
temperature fluctuations to be a probe of ICM turbulence (e.g. Lesieur
1997). However, this relies on the assumption that temperature behaves
as a passive scalar. Once this has been verified empirically, the
almost uniform distribution of the temperature over scales $\gg r_{\rm
c}$ allows a cleaner distinction between small-scale turbulent
substructures and the large-scale cluster profile. However, in reality
temperature maps are affected by cold fronts and other contact
discontinuities which contaminate the diagnostic maps. On the other
hand, pressure maps have a clear relation to velocity ($P=\rho v^2$)
and are not significantly contaminated by contact
discontinuities. Therefore, we regard pressure as a more direct probe
of ICM turbulence.

The mosaic of XMM-Newton observations is well-suited for the detection
of turbulence in the Coma cluster because it allows a better geometric
discrimination between pressure variations originating from the
overall cluster profile, and substructure superimposed onto it. This
transition occurs at about 150\,kpc.

The measured temperature and density gradients (Fig.\,\ref{FIG_GAMMA})
suggest that the substructures have an adiabatic exponent of
$\gamma\,\approx\,4/3$, which is close to the adiabatic case of an
ideal monoatomic gas. On the other hand, contact discontinuities and
strong shocks seem to be less likely in the core region, consistent
with hydrodynamical simulations (Miniati et al. 2000, Miniati
2003). In addition, the statistics of the residual pressure
fluctuations appear quite Gaussian (Fig.\,\ref{FIG_HISTO}) emphasising
their random nature. Their Fourier power spectrum thus completely
summarizes the fluctuating pressure field and can be used to obtain
observational evidence for the presence of turbulent flows which are
characterized by a Kolmogorov/Oboukhov-like spectrum.

Figure\,\ref{FIG_PKFALL} shows the combined power spectrum of the Coma
cluster on scales between 40 and 90\,kpc. A scale-invariant range of
the spectrum is indicated and suggests the detection of an inertial
range of a turbulent ICM. Theoretical three-dimensional power spectra,
\begin{eqnarray}\label{FIT}
{\cal P}_{\rm 3D}(k)\,&=&\,2\pi^2\,k^{-2}\,E_{\rm 3D}(k)\,=\,
2\pi^2\,C_{\rm P}\,\epsilon^{4/3}\,k^{n-2}\,\\
&=&\,2\,\pi^2\,C\,k^{n-2}\,,
\end{eqnarray}
are transformed with Eq.\,(\ref{E30}) into their two-dimensional
counterparts (dashed lines in Fig.\,\ref{FIG_PKFALL}). From the
comparison of observed and theoretical spectra we see that on scales
between 40 and 60\,kpc, the observed power spectrum has a slope
between $n=-7/3$ and $-5/3$. This slope corresponds to the spectral
slope of the Fourier-transformed Kolmogorov/Oboukhov structure
function (Eq.\,\ref{OB1}). On scales between 60 and 90\,kpc the
spectrum bends towards smaller slopes between $n=-5/3$ and $-1/3$. The
normalization constants range from $C=C_{\rm
P}\epsilon^{4/3}=0.0063\,{\rm kpc}^{-4/3}$ for $n=-7/3$, to
$C=0.470\,{\rm kpc}^{2/3}$ for $n=-1/3$. An absolute calibration of
the pressure used in our analyses is in preparation and will give a
direct estimate of $\epsilon$. In this respect it would be interesting
to compare this rate of kinetic energy transport with the observed
X-ray luminosity.

\begin{figure}
\vspace{-0.0cm}
\centerline{\hspace{-0.3cm}
\psfig{figure=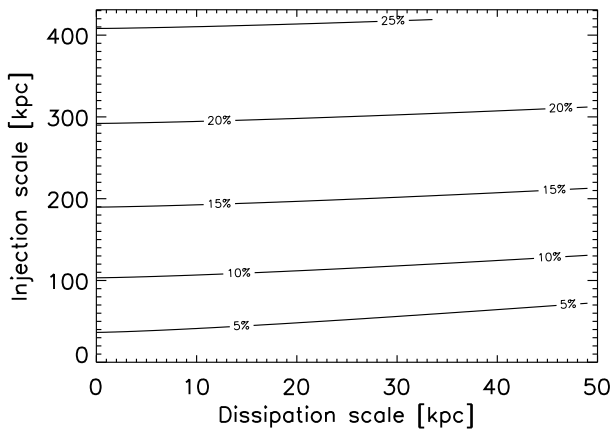,width=9.0cm}}
\vspace{-0.00cm}
\caption{\small Contribution of the turbulent pressure to the total
thermal pressure (contour lines of equal percentage) for a
Kolmogorov/Oboukhov spectrum with the slope $n=-7/3$ and the amplitude
$C=C_{\rm P}\epsilon^{4/3}=0.0063\,{\rm kpc}^{-4/3}$. The power
spectrum is integrated between the injection and the dissipation
scale.}
\label{FIG_CONTRI}
\end{figure}

The power spectrum shown in Fig.\,\ref{FIG_PKF10} allows a first
estimation of the location of the characteristic scale where the
spectrum sharply drops towards larger scales. This scale is at
approximately $\lambda_{\rm i}\,\approx\,100\,{\rm kpc}$ and should be
regarded as a lower limit because of possible contaminations by the
global cluster profile. This scale should also roughly correspond to
the injection scale (e.g., Lesieur 1997), and it is similar to
estimates for the impact parameter of merging clusters based on
kinematics and tidal torque-based arguments (e.g., Sarazin 2002).

The integral of the power spectrum (Eq.\,\ref{E50}) is expected to
give important information about the energy deposited in turbulent
motion. For the scale range between 40 and 90\,kpc, the slope and
amplitude parameters derived above yield relative contributions of the
turbulent pressure to the thermal pressure between 7.4 percent for
$n=-7/3$, and 6.6 percent for $n=-1/3$. The largest possible
contributions are obtained with the $n=-7/3$ spectrum. Therefore, we
computed for this spectrum the relative contribution for different
minimum scales, i.e., lower integration limits of the inertial range
(dissipation scale), and maximum scales, i.e., upper integration
limits (injection scale).

Figure\,\ref{FIG_CONTRI} shows that for a fixed turbulence spectrum
the relative contribution is mainly determined by the value of the
injection scale $\lambda_{\rm i}$. We do not see any turbulent eddies
of the size of the core radius of 420\,pc in Fig.\,\ref{FIG_PWLZ}
which could have erroneously been subtracted by the Fourier low-pass
filter -- although they still might be present, but are difficult to
discriminate from the global cluster profile. Therefore, the relative
contribution of the turbulent pressure to the thermal pressure should
be smaller than 25 percent. If we take the indication for a turnover
in the power spectrum shown in Fig.\,\ref{FIG_PKF10} at $\lambda_{\rm
i}=100\,{\rm kpc}$ as the injection scale, then we would get a lower
limit of about 10 percent. The simulations of Norman \& Bryan (1999)
suggest additional support by turbulent pressure of about 20 percent,
averaged over the cluster (5 to 35 percent between core and virial
radius), which is apparently of the same order as the present
observational limit. However, further study is definitely required in
order to establish how the observational quantities relate to the
simulation results.

For the observed turbulent ICM we can now estimate the kinematic
viscosity by assuming that magnetic fields have a negligible effect
(see below). For a turbulent flow the Reynolds number of the global
fluid $\Re$ measured at the injection scale $\lambda_{\rm i}$, and the
Reynolds number $\Re_{\rm d}$ measured at the dissipation scale
$\lambda_{\rm d}$ are related by $\Re/\Re_{\rm d}=(\lambda_{\rm
i}/\lambda_{\rm d})^{4/3}$. The power spectra do not show any tendency
to decrease at $\lambda=30$\,kpc (Fig.\,\ref{FIG_PKF5}). Therefore,
$\lambda_{\rm d}$ is smaller than 30\,kpc. In the following we assume
a fiducial value of $\lambda_{\rm d}=10$\,kpc. The turbulent flow in
the central region of Coma can thus be characterized by $\Re/\Re_{\rm
d}\,>\,20$. Reynolds numbers at dissipation scales are expected to be
above unity so that $\Re$ will have values in excess of 20. Although
this estimate is rather conservative, this is the best that can be
obtained by direct observations at the moment.

For the viscosity we further need the velocity at the injection
scale. This number can be obtained from hydrodynamical simulations
(Miniati et al., in preparation), which typically give for an 8\,keV
cluster a dispersion turbulent velocity of $v_{\lambda_{\rm
i}}=250\,{\rm km}\,{\rm s}^{-1}$ on scales of $\lambda_{\rm i}\approx
100$\,kpc. This provides a quite reliable upper limit to the kinematic
viscosity of
\begin{equation}\label{NU}
\nu<3\cdot 10^{29}\left(\frac{v_{\lambda_{\rm i}}}{250\,{\rm km}{\rm
s}^{-1}}\right)\left(\frac{\lambda_{\rm i}}{100\,{\rm kpc}}\right)
\left(\frac{\Re}{20}\right)^{-1}\,
\left[\frac{{\rm cm}^2}{s}\right]\,.
\end{equation}
Note that the coherence lengths of magnetic fields in the cores of
galaxy clusters as obtained from Faraday Rotation measurements are
about 5-10\,kpc (e.g., Taylor
\& Perley 1993) and thus below the scale range covered by the present
data. Therefore, we regard the upper limit (Eq.\,\ref{NU}) as not
significantly affected by magnetic fields.

Fabian et al. (2003, see also Reynolds et al. 2004) assume laminar
flow of the ICM around the radio galaxy NGC\,1275 with
$v_\lambda=700\,{\rm km}\,{\rm s}^{-1}$ in the centre of the Perseus
cluster on $\lambda=14\,{\rm kpc}$ scales. From the laminar appearance
of the filaments they assume that the effective Reynolds number is
less than 1000 so that they estimate the lower limit $\nu>4\cdot
10^{27}[\frac{{\rm cm}^2}{s}]$. The upper limit obtained from a
turbulent regime and the lower limit obtained from a laminar regime
can be used to estimate the range 10--30\,kpc where the transition
from a turbulent to a laminar flow could occur. This corresponds to a
dissipation scale of the ICM in the same range. A remark of caution
is, however, necessary here, because we compare two different
situations (merger driven turbulence versus AGN driven turbulence, and
a bulk ICM in Coma versus condensed warm HII-gas in the NGC\,1275
halo), and it is not fully clear in how far they are comparable.

Shibata et al. (2001) determined the 2-point angular correlation
function of hardness ratios as a measure of the temperature
fluctuations detected with ASCA over an area of 19 square degrees in
the Virgo cluster. A significant excess of the correlation amplitude
is found at 300\,kpc. They interpreted the random temperature
fluctuations in Virgo-North as local heating of infalling galaxy
groups.

Future investigations should measure the pressure spectrum of the Coma
cluster more accurately down to 5\,kpc so that the combination with
the present measurements would give information about the ICM in the
Coma cluster from 5--2800\,kpc. This could give us tight constraints
on the type of gas turbulence, its energy content, the importance of
magnetic fields, and on the viscosity of the ICM.

\begin{acknowledgements} 
The XMM-Newton project is supported by the Bundesministerium f\"ur
Bildung und Forschung/Deutsches Zentrum f\"ur Luft- und Raumfahrt
(BMFT/DLR), the Max-Planck Society and the Heidenhain-Stiftung, and
also by PPARC, CEA, CNES, and ASI. We would like to thank Eugene
Churazov and Alex Lazarian for helpful discussions.  We also thank the
anonymous referee for useful comments. AF acknowledges receiving the
Max-Plank-Gesellschaft Fellowship and support from the
Verbundforschung grant 50 OR 0207 of the DLR.  FM was partially
supported by the Research and Training Network ``The Physics of the
Intergalactic Medium'' set up by the European Community under the
contract HPRN-CT2000-00126 RG29185. PS acknowledges support under the
DLR grant No.\,50\,OR\,9708\,35.
\end{acknowledgements}


\begin{thebibliography}{}

\bibitem{} Arnaud, M., Aghanim, N., Gastaud, R., et al. 2001, A\&A, 365, L67

\bibitem{} Batchelor, G.K., 1951, Proc.Camb.Phil.Soc., 47, 359

\bibitem{} Briel, U.G., Henry, J.P., Lumb, D.H., et al. 2001, A\&A, 365, L60 

\bibitem{} Briel, U.G., Finoguenov, A., Henry, J.P. 2003, A\&A submitted

\bibitem{} Burgers, J.M., 1974, The Nonlinear Diffusion Equation
(Dordrecht: Reidel)

\bibitem{} Cho, J., Lazarian, A., 2002, PhRvL, 88, 245001

\bibitem{} Fabian, A.C., Sanders, J.S., Crawford, C.S., Conselic,
C.J., Gallagher III, J.S., \& Wyse, R.F.G., 2003, MNRAS, 344, L48

\bibitem{} Finoguenov, A., Briel, U.G., Henry, P.J., et
  al. 2004a, A\&A, accepted

\bibitem{} Finoguenov, A., Pietsch, W.N., Aschenbach, B.R., Miniati,
F. 2004b, A\&A, 415, 415

\bibitem{} Fritsch, U., Lesieur, M., \& Schertzer, D., 1980, J. Fluid
Mech., 97, 181

\bibitem{} Goldreich, P., \& Sridhar, H., 1995, ApJ, 438, 763

\bibitem{} Henry, J.P., Finoguenov, A., Briel, U.G., 2004, ApJ submitted

\bibitem{} Inogamov, N.A., Sunyaev, R.A., 2003, astro-ph/0310737

\bibitem{} Jansen, J., et al., 2001, A\&A, 365, L1

\bibitem{} Kolmogorov, A.N., 1941, Dokl.Akad.Nauk SSSR, 30, 301

\bibitem{} Lesieur, M., 1997, Turbulence in Fluids, Kluwer Academic
Publisher, Dordrecht

\bibitem{} Lumb, D.H., Finoguenov, A., Saxton, R., et al.
                     2003, SPIE, 4851, 255

\bibitem{} Miniati, F., 2003, NNRAS, 342, 1009

\bibitem{} Miniati, F., Ryu, D., Kang, H., Jones, T.W., Cen, R.,
Ostriker, J.P., 2000, ApJ, 542, 608

\bibitem{} Neumann, D. M., Arnaud, M., Gastaud, R., et al. 2001, A\&A,
365, L74 

\bibitem{} Neumann, D.M., Lumb, D.H., Pratt, G.W., Briel, U.G., 2003,
A\&A, 400, 811

\bibitem{} Norman, M.L., \& Bryan, G.L., 1999, Lect.Not.Phys., 530, 106

\bibitem{} Oboukhov, A.M., 1941, Dokl.Akad.Nauk SSSR, 32, 22

\bibitem{} Oboukhov, A.M., 1949, Dokl.Akad.Nauk SSSR, 66, 17

\bibitem{} Peacock, J.A., 1999, Cosmological Physics, Cambridge
University Press

\bibitem{} Read, A.M. and Ponman, T.J., 2003, A\&A, 409, 395

\bibitem{} Reynolds, C.S., McKernan, B., Fabian, A.C., Stone, J.M., \&
Vernaleo, J.C., 2004, MNRAS (submitted) astro-ph/0402632

\bibitem{} Sarazin, G., 2002, in Merging Processes in Galaxy Clusters,
Eds. L. Feretti, I.M. Gioa, G. Giovannini, APSS Library, Vol. 272,
Kluwer Academic Publishers, Dordrecht, p.\,1

\bibitem{} Shibata, R., Matsushita, K., Yamasaki, N.Y., Ohashi, T.,
Ishida, M., Kikuchi, K., B\"ohringer, H., \& Matsumoto, H., 2001, ApJ,
549, 228

\bibitem{} Taylor, G.B., \& Perley, R.A., 1993, ApJ, 416, 554

\bibitem{} Vestuto, J.G., Ostriker, E.C., \& Stone, J.M., 2003, ApJ,
590, 858

\bibitem{} Vikhlinin, A., McNamara, B.R., Forman, W., Jones, C.,
                     Quintana, H., \& Hornstrup, A. 1998, ApJ, 502, 558

\bibitem{} Vogt, C., En{\ss}lin, T.A., 2003, A\&A, 412, 373

\end{thebibliography}
\end{document}